\documentstyle[epsfig]{article}
        \newcounter{eqnletter}[equation]
\catcode`\@=11 \@addtoreset{equation}{section} \catcode`\@=12

\begin{document}
  {\centerline	{\LARGE {  Sparse Random  Block Matrices}} }
	 \vskip 1cm
{\centerline {	Giovanni M. Cicuta$^*$
\footnote{giovanni.cicuta@gmail.com} and Mario Pernici$^{**}$
\footnote{mario.pernici@mi.infn.it} }} \vskip .3 cm {\centerline  { $^*$
Dept. of Physics, Univ. of Parma, Viale delle Scienze 7A, 43100 Parma,
Italy}} \vskip .3 cm
	{\centerline  {$^{**}$ Istituto Nazionale di Fisica Nucleare,
	Sezione di Milano ,}}
{	\centerline  {	Via Celoria 16, 20133 Milano, Italy}}
		\vskip 1cm {\centerline{Abstract}} \vskip .4 cm

	The spectral moments of ensembles of sparse random block matrices
	are analytically evaluated in the limit of large order.\\ The
	structure of the sparse matrix corresponds to the Erd\"os-Renyi
	random graph. The blocks are i.i.d. random matrices of the
	classical ensembles GOE or GUE.  The moments are evaluated for
	finite or infinite dimension of the blocks.\\
The correspondences between sets of	 closed walks on trees and
classes of irreducible partitions studied in free probability together
with functional
	relations are powerful tools for analytic evaluation of the
	limiting moments. They are helpful to identify probability laws
	for the blocks and limits of the parameters which allow the
	evaluation of all the spectral moments and  of the spectral
	density.\\

\vskip 1 cm
	   \section{Introduction}
		\vskip .6 cm In the past 70 years the theory of random
		matrices had an impressive development in theoretical
		physics and in a variety of disciplines.  It seems
		natural that the study of ensembles of matrices with
		a variety of structures is important to assess the
		possible use of random matrix theory to deal with
		specific problems.\\

		In this paper we study an ensemble of sparse real
		symmetric block matrices. Every matrix $A$ of the
		ensemble has dimension $Nd \times Nd$ and it has $N$
		rows and columns. The $N^2$ entries are  random matrices
		$\alpha_{ij}X_{ij}$, $i,j=1,..,N$, as depicted in
		eq.(\ref{d.3}).
				The set of random variables
				$\{\alpha_{ij}\}$ makes the random block
				matrix sparse, the set of blocks $\{X_{ij}
				\}$  are real symmetric random matrices
				of dimension $d \times d$. Our goal is
				the limiting, $ N \to \infty$, spectral
				distribution of the Adjacency matrix.\\

		Early studies of ensembles of random block matrices
		are \cite{bre}, \cite{gir}. The entries of the
		blocks $X_{ij,\alpha \beta}$,  $i,j=1,..,N$, $\alpha,
		\beta=1,..,d$, are often chosen to be Gaussian random
		variables (possibly with block-dependent variance) and
		the limit of infinite size of the blocks, $d \to \infty$,
		and $N$ finite, is performed. This leads to a system of
		coupled equations for the resolvent, which may be solved
		only in few cases.\\

		More recently, several classes of patterned,
		or structured ensembles of random matrices with
		random blocks were studied.  Some examples  are the
		band matrices or tridiagonal random block \cite{fiod},
		\cite{moli}, \cite{sher}, random block Toeplitz or Hankel
		\cite{Li}, \cite{basu} , block circulant \cite{kologlu},
		structured real symmetric block matrices \cite{oraby1},
		\cite {rogers}, \cite{black}, etc.\\
	Again the limit of infinite size of the blocks, $d \to \infty$,
	is performed.\\

The case  of sparse matrices  is quite different.  Possibly the most
studied model of sparse random matrix ensemble is the Adjacency matrix
of a random graph with $N$ vertices of average vertex degree or average
connectivity $Z$.

\begin{eqnarray} A=\left( \begin{array}{cccccccc} 0 & \alpha_{1,2}
&\alpha_{1,3} & \dots &\alpha_{1,N} \\ \alpha_{2,1}& 0 &\alpha_{2,3}
& \dots & \alpha_{2,N} \\ \dots & \dots & \dots & \dots & \dots\\
\alpha_{N,1} &\alpha_{N,2} & \alpha_{N,3} &\dots & 0 \end{array}\right)
  \qquad , \qquad  \alpha_{i,j}=\alpha_{j,i} \qquad \qquad
\label{d.1} \end{eqnarray}

The set of $N(N-1)/2$ random variables $\{\alpha_{i,j} \}$
, $i>j$, is a set of independent identically distributed
random variables, each one having the probability density
\begin{eqnarray} P(\alpha)=\left(\frac{Z}{N}\right) \delta (\alpha
-1)+\left(1-\frac{Z}{N}\right) \delta(\alpha)\qquad \qquad \label{d.2}
\end{eqnarray}

The random matrix ensemble of eqs. (\ref{d.1}), (\ref{d.2}), was
considered a basic model of disordered system in statistical mechanics. It
was then analyzed for decades from the early days of the replica approach
\cite{pio} up to more recent cavity methods \cite{mod}.\\

More pertinent to this paper, the moments of the spectral density  of
the limiting ($N \to \infty$) Adjacency matrix were carefully studied
and recursion relations for them were obtained \cite{bau}, \cite{khor}.
Remarkably, the knowledge of all the spectral moments, at least in
principle, was not sufficient to obtain the spectral density.\\

In  recent years, an ensemble of sparse random block matrices was
considered, where the entry $A_{i,j}$ of the random matrix  is a real
symmetric $d \times d$ random real symmetric matrix $X_{i,j}$

\begin{eqnarray} A&=&\left( \begin{array}{cccccccc} 0 & \alpha_{1,2}
X_{1,2} &\alpha_{1,3} X_{1,3}& \dots &\alpha_{1,N} X_{1,N}\\
\alpha_{2,1}X_{2,1}& 0 &\alpha_{2,3} X_{2,3} & \dots & \alpha_{2,N}
X_{2,N}\\ \dots & \dots & \dots & \dots & \dots\\ \alpha_{N,1}X_{N,1}
&\alpha_{N,2} X_{N,2}& \alpha_{N,3}X_{N,3}&\dots & 0 \end{array}\right)
  \qquad , \nonumber\\
	&&   X_{i,j}=X_{j,i}, \quad i,j=1, 2,..,N, \nonumber\\
	&&  X_{ij,\alpha \beta}= X_{ij, \beta\alpha }, \quad
	\alpha,\beta=1,2,..,d \qquad \qquad
\label{d.3} \end{eqnarray}

The generic block  $X_{i,j}$ may be considered a matrix weight associated
to the non-oriented edge $(i,j)$ of the graph. One may say that the set of
random variables $\alpha_{i,j} =\alpha_{j,i}$	encodes the architecture
of the non-oriented graph. In this case, see eq.(\ref{d.2}), it is the
Erd\"os-Renyi random graph, with average vertex degree (or connectivity)
$Z$.\\

In the papers  \cite{pari}, \cite{cic1}, \cite{pern}, \cite{benet},
  the
  $d \times d$ real symmetric random matrices $X_{i,j}$ are independent
  (except for the symmetry $X_{i,j}=X_{j,i}$) identically distributed
  projectors with rank one depending on a unit vector $ \vec{n} \in
  R^d$ chosen with uniform probability on the sphere.  This probability
  distribution for the set $\{X_{i,j} \}$ is  unusual in random matrix
  literature. It is motivated by  physics : the random Laplacian block
  matrix associated to the random block matrix $A$ in eq.(\ref{d.3}) is
  the Hessian of a system of points of random locations,  connected by
  springs. \footnote{Actually, in the papers \cite{pari} and \cite{benet},
  the connectivity is fixed, rather than average.}

A surprising result is that it is possible to study analytically the model
in the limit $d \to \infty$, with $t=Z/d$ fixed. The limiting resolvent is
solution of a cubic equation, previously obtained in a different model and
different approximation, the Effective Medium Approximation \cite{semer}.
This random matrix ensemble will be briefly recalled in section 5.1,
where a compact derivation of the cubic equation
of the limiting resolvent follows from a functional transform.\\

In this paper	we analytically evaluate the limiting ($N \to \infty$)
spectral moments  of the Adjacency random block matrix ensemble in
eq.(\ref{d.3}) and the $d \times d$ random matrix blocks $X_{i,j}$
both for finite $d$ and for $d=\infty$. \\

To be definite, for finite $d$,  the random matrix blocks $X_{i,j}$
 are chosen to belong to the Gaussian Orthogonal Ensemble but in the $d
 \to \infty$ limit the results apply to the GUE ensemble as well.  \\
	The motivation of the present work is more theoretical then
	practical. Indeed the sparse random block matrix ensemble of
	this paper has two ingredients : the probability distribution of
	eq.(\ref{d.2}) of the Erd\"os-Renyi random graph and the  Gaussian
	Orthogonal Ensemble of the blocks $X_{i,j}$. Both ingredients
	seem a most natural choice in the random matrix theory.\\

	In section 2 we recall the method of evaluating spectral moments
	as weighted paths on a complete graph. In the present case the
	weight of any path is the trace of the product of the matrices
	associated to the edges. Our analytic result, presented in
	eq.(\ref{A.1}), up to $\mu_{12}$, holds for any probability
	choice for the random matrices $X_{i,j}$ and any dimension $d$. We
	display this rather lengthy equation because it is basic to any
	random block matrix ensemble with the Erd\"os-Renyi  structure
	and the matrix blocks are i.i.d.\\

	The sections 2.2 , 2.3 and 3 recall methods to evaluate  expectations
	of traces of products of random matrices both for $d$ finite and
	for $d\to \infty$.  They follow traditional methods of quantum
	field theory.\\

The limiting moments $\mu_{2k}=\lim_{d \to \infty}\left(\lim_{N \to
\infty} \frac{1}{Nd}<\texttt{tr} A^{2k}> \right)$ are polynomials of
degree $k$ in the  average degree $Z$. They are identical for the choice
of the random blocks $X_{ij}$ belonging to GOE or to GUE.\\ Expectations
of trace of products of random matrices, such as $\lim_{d \to \infty}
<\texttt{tr} (X_1)^{p_1} (X_2)^{p_2}..(X_j)^{p_j}>$, where some of the
blocks are repeated and $p_r$ are positive integers,
 were called multimatrix Gaussian correlators, colored maps, or words
 in several independent Wigner matrices.\\

	Sect.3.1 displays our results for the limiting moments $\mu_{2k}$
	obtained with the above mentioned methods and with the functional
	methods described in sect.4 and 5.\\

	Section 4 uses some tools of free probability and describes
	a generalization of a functional relation, sometimes called
	non-crossing partition transform, or free cumulants. It is indeed
	possible to express the infinite set of weighted paths in which one is
	interested in terms of irreducible weighted paths. This allows the
	evaluation of moments of higher orders and it shows relevant
	relations among different ensembles of random matrices.     \\

 Section 5 makes use of the traditional knowledge  recalled in Sections
 $2$, $3$, and the functional transform of Sect.$4$, to bear on different
 ensembles of sparse block random matrices.\\

There exist probability measures for the matrix blocks $X_{i,j}$ and
limits such that all spectral moments are evaluated and the non-random
spectral density is determined. In other cases the finite number of
spectral moments analytically evaluated is not sufficient to determine
the spectral density, but may be used to develop approximations.\\ In
Sect. 2 we indicate that these different outcomes are related to the
dominant set of closed paths on the graph.

\section{The moments of the Adjacency block matrix}

The method of moments to determine the spectral distribution of random
matrices is one of the oldest and best known techniques, see, for
instance, \cite{kirs},	\cite{greg}, \cite{bai}.  More pertinent to the
present  work are \cite{bau}, \cite{khor}.\\ Every real symmetric, random
or constant, $n \times n$, matrix $m$ corresponds to a non-oriented graph
with $n$ vertices, the edge from vertex $i$ to vertex $j$ corresponds to
the matrix entry $m_{i,j}$. The matrix entry $(m^p)_{v,w}$ corresponds
to the sum of weighted paths of $p$ consecutive steps on the graph from
vertex $v$ to  vertex $w$.\\

Our goal is the determination of the limiting moments $\mu_p$
\begin{eqnarray} \mu_p =\lim_{N \to \infty} \frac{1}{Nd} <  \texttt{tr}
A^{p}>\qquad \qquad \label{d.4} \end{eqnarray} where the matrix $A$
is the block matrix in eq.(\ref{d.3}), the expectation is evaluated
from the probability distribution of the set of $\{\alpha_{i,j}\}$ in
eq.(\ref{d.2}) and the probability distribution of the random blocks
$X_{i,j}$ is discussed later in the paper.\\ Every power $p$ of the
block matrix $(A^p)_{v,v}$ , is a  homogeneous polynomial of degree
$p$ in the set $\{ \alpha_j \}$ and in the matrices $ X_{i,j}$. Since
the weights of the edges are non commutative, some recursive tools are
not available.	It seems useful to consider first the expectation over
the random variables $\{ \alpha_j \}$ and the $N \to \infty$ limit. We
postpone to subsection $2.2$ the evaluation of the expectation of the
trace of  products of $p $ random matrices $X_{i,j}$.\\

Let us call $S_1$ the set
  of closed paths  relevant to the $N \to \infty$ limit of spectral
  moments of the matrix ensemble in eq.(\ref{d.1}). These paths are walks
  on a tree graph, because  walks which include a loop do not contribute
  to the moments, in the $N \to \infty$ limit. In a walk, an edge is
  traversed in one direction then later on is traversed in the opposite
  direction. In the product of matrix blocks which is the weight of the
  walk, the distinct matrix blocks appear an even number of times.
	These paths were enumerated in \cite{bau}, \cite{khor}. The
	powerful recurrence relations may be  generalized to accommodate
	commutative weights for the edges, but it seems difficult to
	accommodate our non-perturbative matrix blocks $X_{i,j}$.\\

The set $S_1$ is the basic set of paths relevant to the moments of sparse
random matrix ensembles. It may happen, for certain probability measures
of the matrix blocks and for certain limits of parameters, that a subset
$S_2 \subset S_1$ of paths is dominant. Each product of matrix blocks in $S_2$
corresponds to a non-crossing partition (as we	explain in Sect.4). The
walks in $S_2$ are enumerated by Narayana polynomials.	This situation
is most interesting because, with different probability measures of
the blocks $X_{i,j}$, it leads to different algebraic equations for
the resolvent.\\

Finally, in the dense limit ($Z \to \infty$), the further subset
$S_3\subset S_2 \subset S_1$  of the Wigner paths is dominant  \cite{molch},
\cite{sinai}, \cite{pastur}, \cite{tao} , \cite{metz}.	In these
paths, there is the maximum number of distinct edges. In the products
of matrix blocks corresponding to a Wigner walk, each distinct block
occurs twice. Wigner walks
 are  in one-to-one correspondence with Dyck paths and are counted by
 the Catalan numbers.  They are well known and we do not discuss the
 set $S_3$ in this paper.\\

\subsection{Closed paths on tree-graphs with weighted edges}

Because of the probability distribution of the $\{ \alpha_{i,j} \}$ random
variables, eq.(\ref{d.2}), in the limit $N \to \infty$, only closed paths
on trees contribute to the limiting moments, \cite{bau}, \cite{khor}.\\
This is true independent from the probability distribution of the random
blocks $X_{i,j}$ and their finite dimension $d$.\\ For a closed  path on
a tree,  every edge is traversed an even number of times. The limiting
moments of odd order vanish.\\ The weight of a path is the product of
the weights of the traversed edges. For a closed walk of $2p$ steps,
the number $l$ of distinct matrices occurring in the weighted path is $1
\leq l \leq p$.\\ Since the blocks are i.i.d., the identification of a
block $X_{i,j}$ in a product is irrelevant.  It is useful a relabeling
of the products of the blocks that only records if the blocks are equal
or different to other  ones in the product. For instance :

\begin{eqnarray} &&X_{1,3}X_{3,1}X_{1,3}X_{3,4}
X_{4,7}X_{7,4}X_{4,3}X_{3,1}\qquad \qquad \texttt{is relabeled} \qquad
\nonumber\\ && (X_1)^3 X_2 (X_3)^2  X_2 X_1 \qquad \qquad \label{d.6}
\end{eqnarray}

As the order of the spectral moments increases, the number of relevant
products also increases. \\ In \cite{pern} we analytically evaluated,
with help of computer symbolic enumeration, the spectral moments
of the ensemble of Adjacency matrix up to $\mu_{26}=\lim_{N \to
\infty}\frac{1}{Nd}<\texttt{Tr}\,A^{26}>$. The ensemble of random block
matrices of ref.\cite{pern} is depicted in eq.(\ref{d.3}).  In	paper
\cite{pern} the blocks $X_{i,j}$ are $d \times d$ rank-one projectors,
whereas in this paper we consider different probability laws, GOE for $d$
finite, GOE or GUE for $d$ infinite.\\ The set of closed walks on trees,
relevant to the spectral moments, does not depend on the probability
distribution of the blocks $X_{ij}$.\\ We list in eq.(\ref{A.1}) in the
Appendix  the relevant products of matrices up to the order $\texttt{Tr}\,
A^{12}$.\\ As a simple check, one may replace every block $X_{i,j}$
with one and obtains the table $1$  of the moments in \cite{bau}.\\

\subsection{The average over the products of random blocks $X_{i,j}$,
for finite $d$}

In this paper we choose the $d \times d$ random matrices $X_{i,j}$ as
independent Gaussian real symmetric matrices. The ensemble of these
random matrices is called Gaussian Orthogonal Ensemble , GOE. The
joint probability distribution of the entries of a random matrix $X$
is described by the  integration measure, see for instance \cite{pand}

\begin{eqnarray} DX&=& \left( \frac{1}{\sqrt{2}}\right)^d
\left(\frac{1}{\sigma \sqrt{2 \pi}}\right)^{\frac{d(d+1)}{2}}
e^{-\frac{\texttt{tr}\,X^2}{4\sigma^2}} \left(\prod_{\alpha>\beta}
dX_{\alpha,\beta} \right)\left(\prod_{\gamma } dX_{\gamma, \gamma}\right)
\quad ,\nonumber\\ \int DX &=& 1 \qquad \qquad \label{d.7} \end{eqnarray}
The off-diagonal entries $X_{\alpha,\beta}=X_{\beta,\alpha}, \alpha \neq
\beta$, are i.i.d. centered normal variables with variance $\sigma^2$. The
$d$ diagonal entries $X_{\gamma,\gamma}$ are i.i.d. centered normal
variables with variance $2 \sigma^2$.\\

 The expectations of products of several matrices are evaluated by Wick
 theorem with the basic Wick contraction (or propagator)

\begin{eqnarray} < X_{\alpha,\beta}
X_{\gamma,\delta}>&=&\left(\delta_{\alpha,\gamma}\delta_{\beta,\delta}+\delta_{\alpha,\delta}\delta_{\beta,\gamma}\right)\sigma^2
\quad , \nonumber\\
 \frac{1}{d}<\texttt{tr} X^2>&=& (d+1)\sigma^2
\qquad \qquad \label{d.8} \end{eqnarray}

Graphical techniques are useful to keep track of the matrix indices. They
were developed and used long ago to evaluate group factors of Feynman
graphs.  \cite{ger},\cite{cvi}, \cite{cann}, \cite{but}. \\

Eq.(\ref{A.2}) displays the evaluation of the averages over the GOE
ensemble of the contributions displayed in eq.(\ref{A.1}).  The evaluation
of monomial powers \begin{eqnarray}
 \frac{1}{d}<\texttt{ tr}\,X^{2p}>  \qquad \qquad
\label{d.8b} \end{eqnarray}

was studied in the past because of its relation with the average resolvent
$g(z)$ $$g(z)=\frac{1}{d} < \texttt{tr} \frac{1}{z I-X}>=\frac{1}{zd}
\sum_{r=0}^\infty < \texttt{tr} \left( \frac{X}{z}\right)^r >$$ Graphical
techniques are efficient and there exist a $5$ terms recursion relation
\cite{ledo} for the monomials of eq.(\ref{d.8b} ) for any finite $d$.\\

We report in Appendix in eq.(\ref{A.2})  the limiting, $N \to \infty$, $d$
finite,  first spectral moments obtained  by evaluating the expectations
of the terms in eq.(\ref{A.1}) by graphical methods. They  provide a check
of the power of a functional transform we describe in  section $4$.\\

Simple consistency  checks may be performed on the evaluations in
eq.(\ref{A.2}).
Let us suppose that  $\{ X_j \}$, $j=1,2,..k$, is a set of i.i.d. random
matrices of the GOE ensemble. Then the random matrix $X=\frac{1}{\sqrt{k}}
\sum_{j=1}^k X_j$ also belongs to the same GOE ensemble, for every $k$
and every $d$. This generates an infinite number of sum rules. For
instance \begin{eqnarray} &&<\texttt{tr}(X_1+X_2+X_3)^6> = <\texttt{tr}
\Bigg[3(X_1)^6+36(X_1)^4(X_2)^2+36(X_1)^3X_2X_1X_2+\nonumber\\
&&+18(X_1)^2X_2(X_1)ìX_2+12(X_1)^2(X_2)^2(X_3)^2+18(X_1)^2X_2(X_3)^2X_2+\nonumber\\
&&+36(X_1)^2X_2X_3X_2X_3+ 6X_1X_2X_3X_1X_2X_3+18 X_1X_2X_3X_1X_3X_2
\Bigg]> \nonumber\\ &&=27< \texttt{tr} (X_1)^6> \label{d.10}
\end{eqnarray}

The $d=1$ case can be computed using recursion relations, which are a small
modification of those in \cite{bau}, as written in \cite{pern2}, for $k=2$;
in these equations there is a sum over $\bar m$, the number of times an 
edge $a$ is run up and down; here there is a Gaussian random variable $X_a$
associated to each time the edge is traversed, so there is a factor 
$X_a^{2\bar m}$;
since in these relations $X_a$ appears only here, one can take the
average 
\begin{equation}
< X_a^{2{\bar m}} > = 2^{\bar m} (2{\bar m}-1)!!
\label{d1av}
\end{equation}
so one has 
\begin{equation}
\mu_{2j} = H_{j,1}(1,Z)
\label{fpsiH}
\end{equation}
where $H_{j,1}(y,Z)$ is given by the recursion relations are
\begin{eqnarray}
&&H_{j, 1}(y, Z) = Z \sum_{\bar m = 1}^j < X^{2{\bar m}} > y^{\bar m}
\sum_{j_1 + j_2 = j - \bar m}
H_{j_1,\bar m}(y,Z)H_{j_2,\bar m}(1,Z) \nonumber\\
\label{psiH1} \\
&&H_{j,\bar m}(y, Z) = \sum_{m=1}^j y^m [y^m]H_{j, 1}(y, Z) \left( \begin{array}{cc} m+\bar m - 1 \\  \bar m - 1 \end{array} \right)
 \label{psiH}
\end{eqnarray}
and $H_{0,\bar m}(y, Z) = 1$.
The moments for generic $d$, which we computed till $\mu_{12}$, agree
in $d=1$ with the moments obtained with these recursion relations.

\subsection{Factorization}

In the product of the random matrices $\{X_j\}$ there can be a subproduct
of a subset of random matrices, none of which appear anywhere else
  in the product. Because of the independence  of the blocks $\{X_j\}$
  with different index $j$, it follows a factorization, which may be
  repeated until the factors cannot be further reduced.
Such factors will be called irreducible . For instance \begin{eqnarray}
&&<\texttt{tr} X_1^2 X_2^2 X_1^2 X_3 X_4^4 X_3 X_2^2>=<\texttt{tr} X_1^2
X_2^2 X_1^2 X_3^2 X_2^2>\frac{1}{d}<\texttt{tr}X_4^4>=\nonumber\\
&&=<\texttt{tr} X_1^2 X_2^2 X_1^2
X_2^2>\frac{1}{d}<\texttt{tr}X_3^2>\frac{1}{d}<\texttt{tr}X_4^4>
\label{d.12} \end{eqnarray} and \begin{eqnarray} <\texttt{tr} X_1 X_2^2
X_1^2 X_3^2 X_4^2 X_3^2 X_4^2 X_2^2 X_1 >=<\texttt{tr}X_1 X_2^2 X_1^2
X_2^2 X_1> \frac{1}{d}<\texttt{tr} X_3^2 X_4^2 X_3^2 X_4^2> \nonumber\\
\label{d.13} \end{eqnarray}

The factorization property holds for arbitrary dimension $d$ and any
probability measure of the random matrices $X_j$.\\ In sect.4 we present
a functional relation which reduces the infinite set of expectations to
the smaller, still infinite set of irreducible expectations.\\

\section{The large-$d$ limit} We now evaluate the first moments of the
spectral density of the Adjacency matrix (\ref{d.3}) in the limit $d\to
\infty$. This is the goal of the present paper.\\

We make the usual choice $\sigma^2 = \frac{1}{d}$ in the Gaussian measure
Eq. (\ref{d.7}).

More precisely, we evaluate the first limiting moments \begin{eqnarray}
 \mu_{2p}&=&\lim_{d \to \infty} \bigg(\lim_{N \to \infty}\frac{1}{N d}
 \texttt{tr}< A^{2p}> \bigg) \quad \nonumber\\
&=& c_1^{(2p)} Z+c_2^{(2p)} Z^2+ \dots +c_p^{(2p)} Z^p\normalfont
\qquad \qquad \label{f.3} \end{eqnarray} where the Adjacency matrix $A$
is given in eq.(\ref{d.3}), the expectations are evaluated over the
set of i.i.d. random variables $\{\alpha_{i,j}\}$, see eq.({\ref{d.2})
which define the sparsity of the block matrix and the GOE measure for
the entries of the i.i.d. blocks $X_{i,j}$ is in eq.(\ref{d.7}).\\

The spectral  distribution  of each GOE block $X$, in the $d \to \infty$
limit, is the semi-circle $$\rho(\lambda)=\frac{1}{2\pi}\sqrt{4-\lambda^2}
\quad , \qquad -2 \leq \lambda \leq 2$$

Then the coefficients $c_1^{(2p)}$ are known for every $p$ to be the
Catalan coefficients $C_p$ \begin{eqnarray} c_1^{(2p)}&=& \lim_{d \to
\infty} \left( \ \frac{1}{d}< \texttt{tr}  X^{2p}>\right)=\nonumber\\
&=&\frac{1}{2\pi}\int_{-2}^2 \lambda^{2p} \sqrt{4-\lambda^2} \,
d\lambda=\frac{1}{p+1} \left(\begin{array}{cc}	2p \\p\end{array} \right)
=C_p \qquad \qquad \label{f.4} \end{eqnarray}

Before discussing further coefficients $c_j^{(2p)}$ . $1\leq j\leq p$,
let us recall methods of evaluation.\\

In the language of quantum field theory, in order to evaluate  a term
like $\lim_{d \to  \infty}\frac{1}{d}<\texttt{tr}[X_1^6X_2^2X_1^2X_2^2]>$,
one would draw a closed convex line, with $12$ points on it, with labels
$X_1$, six consecutive points, then $X_2$, two consecutive points,
etc. The evaluation  of the  expectation is provided by the Wick
theorem and would be called  a Gaussian two matrix expectation. In
this example, there are many distinct ways, or graphs, where six
"`propagators"'  couple points with the same label. Each graph,
where the six propagators are drawn inside  the closed line has a
contribution  to the evaluation for finite $d$. In the limit $d \to
\infty$, the only graphs which contribute are the so called planar
graphs, that is the drawings where the propagators do not cross inside
the border line. Furthermore to evaluate the contribution of a planar
graph, it is sufficient to  retain only one part of the propagator
(\ref{d.8}) \begin{eqnarray} < X_{\alpha,\beta} X_{\gamma,\delta}>&\sim
&\delta_{\alpha,\delta}\delta_{\beta,\gamma}\frac{1}{d}
 \qquad \qquad
\label{f.5} \end{eqnarray}

 This approximate procedure, which is valid to obtain the leading
 contribution in the limit of the large matrices $X$, is identical to the
 procedure for the case of an ensemble of complex self-adjoint matrices
 $X$ belonging to the GUE ensemble. Then the table of limiting moments
 in eq.(\ref{t.1}) holds for both classical ensembles GOE and GUE.\\

Returning to our example, with the above rules, the planar graphs
are only of two disjoint sets : either the two propagators of the two
pairs $X_2^2$ cut across the graph (in only one planar way) or they
couple pairs of adjacent $X_2$. That is

\begin{eqnarray}
 \lim_{d \to  \infty}\frac{1}{d}<\texttt{tr}[X_1^6
 X_2^2 X_1^2 X_2^2]>&=&\lim_{d \to \infty} \left(
 \frac{1}{d^2}<\texttt{tr}[X^6]><\texttt{tr}[X^2]>
+\frac{1}{d}<\texttt{tr}[X^8]>\right)\nonumber\\ &=& 19 \nonumber
\end{eqnarray}

The same argument leads to the functional generator \begin{eqnarray}
&& F(g_1,g_2)=\lim_{d \to  \infty}\frac{1}{d}<\texttt{tr}[
\frac{1}{1-g_1X_1^2}X_2^2\frac{1}{1-g_2X_1^2}X_2^2]>=\nonumber\\
&&\quad =\lim_{d \to  \infty}\frac{1}{d}<\texttt{tr}[
\frac{1}{1-g_1X_1^2}\frac{1}{1-g_2X_1^2}]>+ \lim_{d \to
\infty}\frac{1}{d^2}<\texttt{tr}[ \frac{1}{1-g_1X_1^2}]><\texttt{tr}[
\frac{1}{1-g_2X_1^2}]> \nonumber \end{eqnarray} More general and more
complex recursion relations were obtained by M. Staudacher \cite{staud}.\\

\subsection{The limiting moments}

We present in eq.(\ref{t.1}) the analytic evaluation of the limiting
moments (first $N \to \infty $ then $d \to \infty$) of the rescaled
Adjacency matrix, up to $\mu_{22}$.\\

They were determined for the random blocks $X_j$ being random matrices
of the Gaussian orthogonal ensemble. The table holds as well for $X_j$
being members of the Gaussian unitary ensemble.\\

The moments from $\mu_2$ to $\mu_{12}$ were evaluated after taking the
large $d$ limit of the expectation of the Table 1 of weighted closed
paths on trees in Appendix A, eq.(\ref{A.1}). The moments from $\mu_2$
to $\mu_{22}$ were evaluated by the more powerful methods described in
the next sections.

\vskip 0.4 cm
 $\mu_{2p}=\lim_{d \to \infty} \bigg(\lim_{N \to \infty}\frac{1}{N d}
 \texttt{tr}< A^{2p}> \bigg)$

\vskip 0.6 cm

\begin{eqnarray} \mu_2 &=& Z  \nonumber\\ \mu_4 &=& 2\,Z+2\,Z^2
\nonumber\\ \mu_{6} &=& 5\,Z+12\, Z^2+5\, Z^3 \nonumber\\ \mu_{8}
&=& 14\, Z+62 \,Z^2+ 56\, Z^3+14\, Z^4 \nonumber\\ \mu_{10} &=&
42\, Z+310\,Z^2+465\,Z^3+240\,Z^4+42\,Z^5 \nonumber\\ \mu_{12} &=&
132 \,Z+1542  \,Z^2+3454 \, Z^3  +2816\, Z^4 +990\,Z^5	+132 \,  Z^6
\qquad \nonumber\\ \mu_{14} &=& 429 \, Z+7700\, Z^2+24325 \,Z^3+28182 \,
Z^4+15197\, Z^5 \nonumber\\ &&\quad +4004\, Z^6+429\, Z^7 \nonumber\\
\mu_{16}  &=& 1430\,Z+38726\,Z^2+ 166536\,Z^3+259090\, Z^4+192760\,
Z^5 \nonumber\\ &&\quad +76440\,Z^6+16016\, Z^7+1430\,Z^8 \nonumber\\
\mu_{18} &=& 4862\, Z+196374\, Z^2+1122569\, Z^3+2264256\,Z^4+2197188\,Z^5
\nonumber\\ &&\quad +1178712\, Z^6+366996\, Z^7+63648\, Z^8+4862\,Z^9
\nonumber\\ \mu_{20} &=& 16796\, Z+1004126\, Z^2+7503800\, Z^3+19162220\,
Z^4+23427294\,Z^5 \nonumber\\ &&\quad 16071948 \, Z^6+6676410\,
Z^7+1705440\, Z^8+251940\, Z^9+16796\,Z^{10}\nonumber\\ \mu_{22} &=&
58786 \, Z+5175874\, Z^2+49956456 \, Z^3+158795516 \, Z^4 \nonumber\\
&&\quad +238949832\, Z^5+202538688 \, Z^6+106069733\, Z^7+35787906\,
Z^8\nonumber\\ &&\quad +7738434 \, Z^9+994840\, Z^{10}+58786\, Z^{11}
\nonumber\\ \label{t.1} \end{eqnarray}

\section{Non-crossing partitions, irreducible partitions,  the
non-crossing transform. }

The role of the combinatorics of non-crossing partitions (definitions here
below), functional transforms for the spectral measures of some random
matrices ensembles in the large dimension limit were studied  in a series
of papers, see for instance \cite{bani}, \cite{nica}, \cite{mingo}.\\

In this section we describe these methods as suited for this paper but
also provide a non-crossing transform at the level of product of matrices,
more general than the more usual transform for expectations. This
generality is useful to see relations among different but related matrix
ensembles.\\

We begin by recalling some definitions.\\

\subsection{Basic definitions }
Let \textsl{S} be an ordered  set. Then $\pi =\{ B_1,..,B_p\}$ is
a partition of \textsl{S} if the $B_i \neq \emptyset $ are ordered
and disjoint sets, called blocks of the partition, whose union is
\textsl{S}.\\ In this paper the elements of \textsl{S} are  real
symmetric $d \times d$ matrices $X$, the blocks of the sparse random
block matrix ensemble (\ref{d.3}).  The contribution  of a path of length
$2l$ on a tree graph  is the product of $2l$ matrices. For instance  the
product of eight matrices, contributing to the asymptotic moment $\mu_8$,
$X_1 X_2^2 X_1^2 X_2^2 X_1$ , corresponds to the partition in two blocks
$\pi=\{1,\,4,\,5,\,8\} \{2,\,3,\,6,\,7\}$.\\ Elements of \textsl{S} may be
drawn as points on a oriented line, labeled with the block they belong.
A partition is said to be \textsl{irreducible} if there is no set of
points, all belonging to the same  block, which forms a proper interval
on the line. The above partition $\pi$ of a set of $8$ elements into
two blocks is irreducible.\\ A partition  with a block such that its
elements are a set of consecutive points is \textsl{reducible}. The
corresponding product of matrices $\{X_j\}$ contains a monomial factor
of a matrix that does not appear elsewhere in the product. In this case,
as it was described in section 2.3, the expectation of the trace of the
product of matrices factorizes in two traces of products and the result
may further be reduced. Then a factorizable partition  may be reduced to
a product of irreducible partitions.\\ To any product of $2p$ matrices we
may associate the set of $2p$ products obtained by a cyclic permutation of
the matrices. We call this set the \textsl{orbit} of anyone product. The
size of the orbit is the number of distinct products in the orbit,
after relabeling the matrices.\\ For instance the orbit of the product
$X_1^8 X_2^2 X_3^2$ has size 12, the orbit of $X_1^4 X_2^4 X_1^4 X_3^4$
has size $8$, the orbit of $X_1^{12}$ has size $1$.\\

Let us define the insertion of a partition in a partition. Let us consider
the product of $8$ matrices  $X_1 X_2^2 X_1^2 X_2^2 X_1$
corresponding to  $\pi=\{1,\,4,\,5,\,8\} \{2,\,3,\,6,\,7\}$. The expanded
form of the product  $X_1 X_2 X_2 X_1 X_1 X_2 X_2 X_1$ corresponds to
the sequential form $\{1,2,2,1,1,2,2,1\}$.\\ We now want to insert
the monomial $M$ corresponding to a irreducible partition, e.g.
$M=X^2$ in the case of $\{1,2\}$, inside
this product. We denote such insertion $I(M)$. Then
the label of all matrices after that place remains unchanged if the
matrix also appear in positions before	the insertion but are shifted
by the number of integers of the distinct matrices of the insertion,
otherwise. The labels of the matrices of the insertion are shifted by
the number of distinct labels before the insertion.
 For instance
\begin{eqnarray} X_1 I(X^2) X_2^2 X_1^2 X_2^2 X_1 &=& X_1^2 X_2^2 X_3^2
X_1^2 X_3^2 X_1 \nonumber\\ X_1 X_2^2 X_1^2 I(X^2) X_2^2 X_1  &=& X_1
X_2^2 X_1^2 X_3^2 X_2^2 X_1 \nonumber\\ X_1 X_2^2 X_1^2 I(X_1^2 X_2^2
X_1^2 X_2^2) X_2^2 X_1&=& X_1 X_2^2 X_1^2 X_3^2 X_4^2 X_3^2 X_4^2 X_2^2
X_1 \qquad \nonumber\\ \label{t.2} \end{eqnarray}

\subsection{The functional transform}

We  define $A(X)$ as the sum (finite or infinite) of all irreducible
products of  $X_j$ matrices corresponding to possible closed paths, and
define $F(X)$ the sum of all  products of  $X_j$ matrices corresponding
to possible closed paths.
(The notation might be a bit confusing, we have used the symbol $A$
for the Adjacency block matrix; here it is the sum of products of $X$'s,
which are considered to be noncommuting variables).
Following \cite{cal} and \cite{pern2}, one can obtain all partitions
inserting all partitions between consecutive elements of the
irreducible partitions, or before their first element.
The insertion operator allows to obtain
in a recursive way the full sum $F(X)$ from the irreducible products
\footnote{The reader familiar with Feynman graphs will notice the analogy
of eq.(\ref{t.3}) to the Dyson equation expressing the two point function
in terms of the self-energy, the Bethe Salpeter equation expressing the
four-point function in terms of a two-particle irreducible kernel and
more generally the relation of connected Green functions to skeleton
expansions \cite{sym}.} 

\begin{eqnarray} F(X)=1+A\left( I(F) X\right)
\label{t.3} \end{eqnarray}

In this equation\footnote{On the right side of the equation, the number
$1$ stands for the Identity matrix. We reserved the  notation I for
the insertion operator} every matrix $X$ in the sum of the irreducible
products in $A(X)$ is replaced by $I(F)X$. The equation can be used
to construct all the partitions with $p$ elements from the irreducible
partitions of order less or equal to $p$.\\

Let us illustrate how one obtains $F$ till order $p=6$, from the
irreducible terms $A=X_1^2+X_1^4+X_1^6$.  At order $2$, $X_1^2 \to
I(F)X_1 I(F)X_1=X_1^2$.\\ So $F=1+X_1^2$ at order $2$.\\ At order $4$,
the replacements

\begin{eqnarray} X_1^2 &\to & I(F)X_1 I(F)X_1=\nonumber\\ &&I(X^2)X_1
I(1)X_1 +I(1)X_1 I(X^2)X_1=\nonumber\\ &&X_1^2 X_2^2+X_1 X_2^2 X_1
\nonumber\\
 X_1^4 &\to & I(1)X_1 I(1)X_1 I(1)X_1 I(1)X_1 =X_1^4
\nonumber \end{eqnarray} \noindent So $F=1+X_1^2+X_1^2 X_2^2+X_1 X_2^2
X_1+X_1^4=1+F_2+F_4$ including order $4$.\\ Next, at order $6$, the
replacements

\begin{eqnarray} X_1^2 &\to & I(F)X_1 I(F)X_1=\nonumber\\ && I(F_4)X_1
I(1)X_1+I(1)X_1 I(F_4)X_1+I(F_2)X_1 I(F_2)X_1\, ; \qquad\nonumber\\
&&I(F_4)X_1^2=I\left(X_1^2 X_2^2+X_1 X_2^2 X_1+X_1^4\right)
X_1^2=\nonumber\\ &&\quad = X_1^2 X_2^2 X_3^2+X_1 X_2^2 X_1 X_3^2+X_1^4
X_2^2 \, ;\nonumber\\ && X_1 I(F_4)X_1=X_1 I\left(X_1^2 X_2^2+X_1 X_2^2
X_1+X_1^4\right) X_1=\nonumber\\ && \quad = X_1 X_2^2 X_3^2 X_1+X_1
X_2 X_3^2 X_2 X_1+ X_1 X_2^4 X_1\, ;\qquad\nonumber\\ &&I(F_2)X_1
I(F_2)X_1=I(X_1^2)X_1 I(X_1^2) X_1=\nonumber\\ &&\quad =X_1^2 X_2
I(X_1^2)X_2=X_1^2 X_2 X_3^2 X_2 \nonumber \end{eqnarray}

\begin{eqnarray} X_1^4 &\to & I(F)X_1 I(F)X_1I(F)X_1 I(F)X_1=\nonumber\\ &
=& I(F_2)X_1^4+ X_1 I(F_2)X_1^3+ X_1^2I(F_2)X_1^2 + X_1^3 I(F_2)X_1=\qquad
\nonumber\\ &=& X_1^2 X_2^4+ X_1 X_2^2 X_1^3+X_1^2 X_2^2X_1^2+X_1^3
X_2^2 X_1 \nonumber \end{eqnarray}

\begin{eqnarray} X_1^6 \to  I(F)X_1 I(F)X_1I(F)X_1
I(F)X_1I(F)X_1I(F)X_1=X_1^6\qquad \nonumber \end{eqnarray}

$F_6$ is the sum of the $12$ terms here obtained. After
taking the trace, using the cyclic property and relabeling
one has \begin{eqnarray} \texttt{tr} F_6=\texttt{tr}X_1^6 +6\,
\texttt{tr} \left( X_1^4 X_2^2 \right)+2\, \texttt{tr }(X_1^2 X_2^2
X_3^2)+3\,\texttt{tr}(X_1^2X_2X_3^2X_2) \qquad \nonumber \end{eqnarray}
in agreement with the evaluation of $\texttt{tr}A^6$ in the third line
of eq.(\ref{A.1}).\\

Furthermore, after expanding the polynomials and  by keeping terms
up to products of six  matrices, one may check that \begin{eqnarray}
\frac{1}{d}< \texttt{tr} F>=1+\frac{1}{d}<\texttt{tr}\Bigg(X_1^2
\left(\frac{<\texttt{tr}F>}{d}\right)^2+X_1^4\left(\frac{
<\texttt{tr}F>}{d}\right)^4+X_1^6\left(\frac{
<\texttt{tr}F>}{d}\right)^6\Bigg)> \nonumber \end{eqnarray}

The reason of this rearrangement is that the matrices inserted with $I(F)$
in all factors of product of matrices have labels different from any
other factor of the irreducible partition, therefore  factorize in the
expectation.\\ Equation (\ref{t.3}) then implies the simpler equation
\begin{eqnarray} \frac{1}{d}< \texttt{tr} F>=1+\frac{1}{d}<\texttt{tr}
A\left( X \frac{<\texttt{tr}F>}{d} \right)> \label{nct0} \end{eqnarray}

Let us multiply each block $X_j$ by the variable $x$, then,
in this example, $A(x X)=x^2 X_1^2+x^4 X_1^4+x^6 X_1^6$,
$$a(x):=\frac{1}{d}<\texttt{tr}A(x X)>=x^2\frac{1}{d}<\texttt{tr}X^2>+
x^4\frac{1}{d} <\texttt{tr}X^4>+x^6\frac{1}{d}<\texttt{tr}X^6> $$
$f(x):=\frac{1}{d}<\texttt{tr}F(x X)>$ solves the simpler equation

\begin{eqnarray} f(x)=1+a\left(x f(x) \right) \label{t.5} \end{eqnarray}

This equation is known	as non-crossing partition transform \cite{bei},
\cite{cal}, or it is called the relation between spectral moments and free
cumulants \cite{spei}.\\ It is also useful to introduce a parameter $p$,
counting the number of blocks of the partition.\\

$A(x X,p)$ is the sum of all irreducible products of $X_j$ matrices
corresponding to possible closed paths on trees. Each product in the
sum has the factor $x^{2l} p^h$ if the product has $2l$ matrices and $h$
are distinct.\\

\begin{eqnarray} a(x,p)&:=&\frac{1}{d}<\texttt{tr}A(x X,p)> \quad , \quad
    f(x,p):=\frac{1}{d}<\texttt{tr}F(x X,p)> \label{t.6a}\\ f(x,p)&=&1+a\left(x
f(x,p),p \right) \label{t.6} \end{eqnarray}

In the next section, these simple functional equation will be used to
obtain the moments of a limiting spectral density for different matrix
ensembles.\\ More remarkably, the resulting different models only depend
upon the same set of irreducible partitions, the set of closed paths
on trees and the probability density to evaluate the expectations.\\ It
is then useful to recall the irreducible products, from eq.(\ref{A.1})
up to the products of $12$ matrices; each of these monomials corresponds
to a representative partition of an orbit; the integer coefficient 
is the size of the orbit.
  \begin{eqnarray}
&& p\Bigg( X_1^2+X_1^4+X_1^6+X_1^8+X_1^{10}+X_1^{12} \Bigg) \nonumber\\ &&
p^2 \Bigg( 2\, X_1^2 X_2^2 X_1^2 X_2^2 +10\, X_1^4 X_2^2 X_1^2 X_2^2+12\,
X_1^6 X_2^2 X_1^2 X_2^2+12\, X_1^4 X_2^4 X_1^2 X_2^2+\nonumber\\ && \qquad
+6\,X_1^4 X_2^2 X_1^4 X_2^2+2X_1^2 X_2^2 X_1^2 X_2^2 X_1^2 X_2^2\Bigg)
\nonumber\\ && p^3 \Bigg(6\, X_1^2 X_2^2 X_1^2X_3^2 X_2^2 X_3^2+3\,X_1^2
X_2 X_3^2 X_2 X_1^2 X_2 X_3^2 X_2+2 \, X_1^2 X_2^2 X_3^2 X_1^2 X_2^2
X_3^3 \Bigg) \nonumber\\ \label{t.7} \end{eqnarray}

We recall that the $2l$-th moment of the Adjacency matrix of
eq.(\ref{d.1})-(\ref{d.3}), in the $N \to \infty$ limit, is given by
the set of closed walks on tree graphs, associating a factor $Z=p$
for each distinct edge of a walk.
In ref.\cite{pern2} such set
is proved to be isomorphic to $P_2^{(2)}(l)$, the set of $2$-divisible
partitions with $2l$ elements, with the restriction that, given two
distinct blocks $B_i$ and $B_j$ of a partition, there is an even number of
elements of $B_j$ between any pair of elements of $B_i$.
This isomorphism has been found first in \cite{bose}, in which this
set of partitions, defined in an equivalent way, is called $SS(2l)$.
\\ Furthermore,
the set $P_2^{(2)}(l)$ has the property that it can be factorized in
irreducible partition under non-crossing. Then the generating function
$f$ of the number of partitions with given number of elements is related
to the generating function $a$ of the number of irreducible partitions
with given number of elements \cite{bei}.\\

It may happen that the generating function $a(x,p)$ of the irreducible
partitions is not known in closed form, but only its truncated
Taylor expansion.  The eq.(\ref{t.6}) is then used to evaluate the
coefficients of the Taylor expansion of the generating function  $f(x,p)$.
\begin{eqnarray} a(x,p)&:=& \sum_{n \geq 1} a_n(p) x^n\quad , \quad
f(x,p):= 1+\sum_{n \geq 1}f_n(p) x^n \nonumber\\ \sum_{n \geq 1}f_n(p)
x^n &=& \sum_{n \geq 1}  a_n(p) x^n \left(1+\sum_{m \geq 1}f_m(p) x^m
\right)^n \label{t.8} \end{eqnarray}

By comparing the same power of $x$, one obtains the coefficients $f_n(p)$
as functions of the $a_j(p)$ and viceversa. We display in Table 3 the
lowest order relations.\\

\section{Different and related	matrix ensembles}

Depending on the probability law of the random $d \times d$ random blocks
$X_{i,j}$, the functional equations of the previous section may lead to an
algebraic equation for the resolvent of the matrix ensemble or they are
only useful to express the spectral moments in terms of free cumulants.
In all cases they show similarities and differences among different
matrix ensembles.\\

\subsection{The elastic springs, matrix block $X_{i,j}$ is a rank-one
projector}

As it was mentioned in the Introduction, a sparse random block matrix
ensemble was proposed for the study of vibrations of amorphous solids
\cite{pari}, \cite{cic1}, \cite{pern}, \cite{benet}. The
ensemble of Adjacent matrices has the form of eq.(\ref{d.3}) , the
variables $\alpha_{i,j}=\alpha_{j,1}=$ are i.i.d. random variables
with the probability law  in eq.({\ref{d.2}), the $d \times d $ blocks
$X_{i.j}$ are rank-one projectors $$X_{i,j}=X_{i,j}^{t}=X_{j,i}=\hat
n_{ij} \hat n_{ij}^t $$ where  $\hat n_{ij}$ is a $d$-dimensional
random vector of unit length, chosen with uniform probability on the
$d$-dimensional sphere and $\hat n_{ij} \hat n_{ij}^t $ is the usual
matrix product of a column vector  times a row vector originating a
rank-one matrix.\\

It was proved in \cite{pern} that the equations for the generating
functions of the spectral moments  are greatly simplified in the $d \to
\infty$ limit, with $t=Z/d$ fixed, then obtaining a cubic equation for
the limiting resolvent. The proof in \cite{pern} allows the use of the
equation (\ref{t.6}) to obtain the limiting resolvent in few lines. Indeed
in this limit, out of the whole set $S_1$ of the closed paths on trees,
relevant for finite $d$,  only the subset $S_2 \subset S_1$ survive,
each one with weight one.\\

The evaluation of the limiting moments becomes the counting
of non-crossing partitions. The sum  $A(X,Z)$  of all irreducible
products of  $X_j$ matrices corresponding to possible closed paths $
A(X,Z)=Z\sum_{l \geq 1} (X_1)^{2l}$. 

It is easy to see also in this model from Eq. (\ref{t.3}) one
obtains Eq. (\ref{nct0}), where $F(X, Z)$ depends on $Z$, counting the number
of blocks, and on $d$; as shown in \cite{pern}, taking the average
there is a factor $\frac{1}{d}$ for each block of the partition
(i.e. for each distinct matrix block $X$), so that the dependence
of $F$ on $Z$ and $d$ is through $t=\frac{Z}{d}$. Therefore in
Eq.(\ref{t.6}) one has $p=t=\frac{Z}{d}$.

\begin{eqnarray} a(x,p)&=& p\sum_{l \geq 1}x^{2l}=p \frac{x^2}{1-x^2}
\nonumber\\ f(x,p) &=& 1+p \frac{ x^2 f(x,p)^2}{1- x^2 f(x,p)^2}
\label{t.10} \end{eqnarray}

This cubic equation  was the result of reference \cite{pern} for the
model of random  elastic springs, in the limit $d \to \infty$ with
$p=Z/d$ fixed.\\ Very recently A. Dembczak-Kolodziejczik and A. Lytova,
by methods of the resolvent, provided an alternative proof of this result
and the analogous one for the Laplacian ensembles \cite{aa}.\\

The function $f(x,p)=\sum_{n=0}^\infty x^{2n}\mu_{2n}$ is a generating
function of the moments of the spectral density. The moments are
Fuss-Narayana polynomials $$ \mu_{2n}=\sum_{b=1}^n \frac{1}{b} \left(
\begin{array}{cc} n-1\\b-1 \end{array}\right) \left( \begin{array}{cc}
2 n\\b-1 \end{array}\right) \,p^b$$

The equivalent cubic equation for the resolvent was obtained by
Semerjian and Cugliandolo long ago \cite{semer} as an approximation,
called the Effective Medium Approximation, on the replica formalism for
a \textsl{different} matrix ensemble. It was also obtained by Slanina
with the cavity method \cite{sla}.  Very recently  different methods of
obtaining the spectral density ensembles of sparse real symmetric random
matrices have been reviewed \cite{vivo}.\\

The spectral density defined by the generating function $f(x,p)$ is a
member, $\pi_{2,t}$, of a remarkable family of measures, called Free
Bessel laws \cite{bani}.  The authors discussed the combinatorial
relevance of these measures and the relation of $\pi_{2,t}$ to
non-crossing partitions where all blocks have an even number of
elements. It was not shown a random matrix ensemble related to this
spectral density.\\

\vskip 2cm \subsection{Matrix block $X_{i,j}$ is element  of GOE ensemble,
$d$ finite}

 In sections 2.2 and 2.3 we summarized rules to evaluate  expectations of
 products of block matrices $X_j \in$ GOE, for finite $d$. The products
 in which we are interested are the terms in eq.(\ref{A.1}) contributing to the
 spectral moments in the limit $N \to \infty$. The spectral moments of
 low order are reported in eq.(\ref{A.2}).\\

Here we test the usefulness of the functional transform in a case where
the generating function of irreducible partitions $a(x,p)$ is not known
in closed form, but merely its truncated Taylor expansion.\\

Up to the product of $12$ block matrices, the irreducible terms
are listed in eq.(\ref{t.7}). The expectation of the trace
of the monomial powers are evaluated by the recursion relation
\cite{ledo} \begin{eqnarray} &&\frac{p}{d} <\texttt{tr} \Bigg(x^2
X_1^2+x^4X_1^4+x^6X_1^6+x^8X_1^8+x^{10}X_1^{10}+x^{12}X_1^{12}
\Bigg)> = \nonumber\\ &&p\,\Bigg( \sigma^2 x^2(d+1)+\sigma^4 x^4
(2d^2+5d+5)+\sigma^6 x^6 (5d^3+22d^2+52d+41)+\nonumber\\ &&\sigma^8 x^8
(14 d^4+93 d^3+374 d^2+690 d+509)+\nonumber\\ && \sigma^{10}x^{10}
(42 d^5+386d^4+2290d^3+7150 d^2+12143 d+ 8229)+ \nonumber\\ &&
\sigma^{12}x^{12} (132 d^6+1586 d^5+12798 d^4+58760 d^3+167148 d^2+258479
d+166377) \Bigg) \nonumber \end{eqnarray}

Other terms are reduced to evaluation of pure powers, by Wick
contractions of the  $X_2$ fields. For instance \begin{eqnarray}
&&\frac{1}{d}<\texttt{tr} (X_1)^{2n} X_2^2 X_1^2 X_2^2>= \nonumber\\
&& \sigma^4 \Bigg( (d^2+3d+5)<\texttt{tr} (X)^{2n+2}>+\sigma^2
(d+2)(d^2+d+4n) <\texttt{tr} (X)^{2n}>\Bigg)
 \nonumber
\end{eqnarray}

 The remaining irreducible terms, up to order $12$ are also evaluated
 by graphic methods. It is easy to use the evaluated irreducible terms
 $a_j$ in the series (\ref{t.8}) and Table $4$ to evaluate the moments and
 verify Table $2$ or extend it to higher orders. Indeed, in this ensemble,
 both free cumulants and spectral moments of odd order are zero. Then,
 for example
$$f_8=a_8+8 a_2 a_6+4 a_4^2 +28 a_2^2 a_4+14 a_2^4 $$ This subsection
shows the the functional transform may be convenient even for finite
$d$.\\

\vskip 2cm \subsection{Matrix block $X_{i,j}$ is element  of GOE or GUE
ensemble, $d$ infinite}

To evaluate the spectral moments in the limit $d \to \infty$ one chooses the
variance $\sigma^2=1/d$ in eq.(\ref{d.8}).\\ As is well known, the
leading order expectations is the same for random matrices in the GOE
or in the GUE ensembles.\\

Expectations of trace of products of random matrices in the $d \to
\infty $ limit are simpler than for finite $d$.\\ We report here our
evaluation of the generating function of irreducible terms, $a(x,Z)$, in
the $d \to \infty $ limit, up to order $22$.  \begin{eqnarray} a(x,Z)&=&
Z x^2 + 2\, Z x^4 + 5\, Z x^6 + (14 \,Z+6\, Z^2) x^8 +(42\, Z+70 \,Z^2)
x^{10}+ \nonumber\\
 &&(132\,Z+552 \, Z^2+50\, Z^3) x^{12}+(429 \,Z+3696 Z^2+1204 \,Z^3)
 x^{14}+ \nonumber\\
&& (1430\, Z+22710\, Z^2+17272\, Z^3+618 Z^4) x^{16}+\nonumber\\
&&(4862\,Z+132726\, Z^2+193289 Z^3+23808 Z^4) x^{18}+\nonumber\\ &&
(16796\,Z+752186\,Z^2+1869200\,Z^3+518600\,Z^4+9606\,Z^5)x^{20}+\nonumber\\
&&(58786 Z+4181034\,Z^2+16446826 \,Z^3+8459308\,Z^4+ 524040\, Z^5)x^{22}
\nonumber \end{eqnarray}

 This truncated generating function and the perturbative functional
 transform lead to  the spectral moments up to $\mu_{22}$, reported
 in eq.(\ref{t.1}).
The values of the moments $\mu_2$ up to $\mu_{12}$ were independently
computed from the table of moments in the Appendix, eq.(\ref{A.1}).\\

It is interesting to compare these moments with a work by Z.D. Bai,
B. Miao, B. Jin , \cite{bai2007}. The authors study an ensemble of random
matrices, each being the product of two random matrices, $S_n W_n$, where
$S_n=\frac{1}{n}  X_n X_n^{\dag}$, $X_n$ is a rectangular $p \times n$
matrix with complex i.i.d. entries, $W_n$ is a Wigner matrix.\\ The
authors prove that  in the limit $n \to \infty, p \to \infty$,	$p/n=y$
finite, the spectral distribution of the  random ensemble  $S_n W_n$
converges to a non-random   density function, eq.(1.6) of their paper,
and the spectral moments converge to the non-random moments $\beta_k$
we reproduced in Table 4 in the Appendix.\\

Let us show that $Z \beta_k(y=Z)$ is equal to an approximation of the moment
$\mu_k$ of the Adjacency block matrix model in the limit $d\to \infty$,
considered in this paragraph, where the approximation consists in
considering only walks whose sequence of steps  correspond to 
noncrossing partitions, i.e. considering only partitions belonging to
$NC^{(2)}(h)$, instead of those in $P^{(2)}_2(h)$.
The generating function of the irreducible terms of the latter model
is the series of the power monomials,
and the expectations are the limiting,  $d \to \infty$, values for GOE
or GUE ensembles. Let us use $p = Z$ in the following.

\begin{eqnarray} a(x,p)&=& \lim_{d \to \infty} \sum_{k \geq
1}\frac{p}{d}<\texttt{tr} X^{2k}> x^{2k} = p\sum_{k \geq 1} x^{2k}
\frac{1}{k+1}\left( \begin{array}{cc} 2k\\k \end{array}\right)=\nonumber\\
&=& p\left(\frac{ 1-\sqrt{1-4x^2}}{2 x^2 }-1\right) \nonumber
\end{eqnarray}

This closed form may be used in eq.(\ref{t.6}) to obtain the
generating function $f(x,p)$ of the spectral moments \begin{eqnarray}
&&f(x,p)=1+p\left( \frac{1-\sqrt{1-4 x^2 f^2(x,p)}}{2 x^2 f^2(x,p)}-1
\right) \qquad , \qquad \texttt{or}\nonumber\\ && x^2 f^2(x,p) \bigg(
f(x,p)+p-1 \bigg)^2-p f(x,p)+p=0 \qquad \qquad \label{t.30} \end{eqnarray}

Changing variables $f = 1 + p z g$, one gets
\begin{eqnarray}
&&g(z) = {\cal A}(z g(z)) \nonumber \\
&&{\cal A}(z) = (1 + p z)^2 (1 + z)^2
\end{eqnarray}
where $z = x^2$.
From Lagrange inversion formula,
\begin{eqnarray}
\beta_{2n} = [z^{n-1}] g(z) &=& \frac{1}{n}[z^{n-1}] {\cal A}(z)^n = 
\frac{1}{n}[z^{n-1}] ((1 + p z)^{2n} (1 + z)^{2n}) \nonumber\\
&=&\frac{1}{n}\sum_{i=0}^{n-1} p^i \left( \begin{array}{cc} 2n \\i  \end{array}\right)\left(\begin{array}{cc} 2n \\n-1-i  \end{array}\right)
\nonumber
 \end{eqnarray}
With $k=2n$, $y=p$ and $i=j+1$ one obtains Eq. (\ref{A.6}).\\

The generating function $f(x,p)$ of the spectral moments is a
solution of the quartic  equation (\ref{t.30}). We reproduce in
Table $4$ the spectral moments and compare them with  the ones in
eq.(\ref{t.1}). In the latter case, the spectral moment $\mu_{2l}$ is a
polynomial $\mu_{2l}=\sum_{k=1}^l b_k^{(2l)}Z^k$, the lowest coefficient
$b_1^{(2l)}$ and the two highest ones $b_{l-1}^{(2l)}$, $b_{2l}^{(2l)}$
are the same of the corresponding ones in Table $3$ whereas  the remaining
$b_k^{(2l)}$ are bigger  integers than the analogous ones in Table $4$
because  the latter ones have contributions only from closed paths
in $S_2$.

\vskip 2cm \section{Conclusions}

The moments of spectral density of sparse random block matrices, with the
architecture of the Erd\"os-Renyi random graph, are analytically studied
in two steps. In the first step we list the products of blocks pertinent
to the limiting spectral moments, up to $\mu_{12}$ for any probability
distribution of the i.i.d. blocks $X_{i,j}$, random matrices of finite or
infinite dimensions. In the second step, we evaluate the expectations of
the products, for several probability laws of the i.i.d. blocks $X_{i,j}$,
GOE for $d$ finite, GOE or GUE for $d$ infinite.\\ With limited effort,
our list may be adapted to the evaluation of sparse random block matrices
with the architecture of a random regular graph.\\

Three sets of set partitions $S_j$ , $j=1,2,3$, defined in Sect.2,
correspond to sets of closed walks on trees of the weighted graph
which add up to the spectral moments. The relevant set $S_j$ depends
on the probability law chosen for the blocks and further limits of the
parameters.\\

We do not study here the well known set $S_3$ , $S_3 \subset S_2 \subset
S_1$, of the Wigner closed paths.  Two examples are provided where the
relevant set is $S_2$. It corresponds to the non-crossing partitions. In
this case, it is possible to evaluate all the moments and the spectral
density, with help of the functional transform.\\

For other important probability laws of the blocks and limiting values of
the parameters, the moments are evaluated by the largest set $S_1$. Then
it is possible to evaluate the spectral moments at high order, but we
know no example where the spectral density was evaluated.\\

\vskip 3cm \section{Appendix} 

 Table 1. The weighted closed paths on trees relevant to the limiting
 moments (that is $N \to \infty $) of the Adjacency matrix.
\vskip 0.8 cm

 \footnotesize
\begin{eqnarray}
  \frac{1}{N} tr\,A^2 &=& Z\,tr\,(X_1)^2  \qquad \qquad \qquad
  \nonumber\\
\frac{1}{N} tr\,A^4 &=& Z\,tr\,(X_1)^4+Z^2\,2\,tr\,(X_1)^2(X_2)^2 \qquad
\qquad \qquad
 \nonumber\\
\frac{1}{N} tr\,A^6 &=& Z\,tr\,(X_1)^6+Z^2\,6\,tr(X_1)^4(X_2)^2
+ Z^3\,tr\,\left[ 2\,(X_1)^2(X_2)^2(X_3)^2 +3\,(X_1)^2
X_2(X_3)^2X_2 \right] \qquad \nonumber\\ \frac{1}{N}
tr\,A^8 &=& Z\,tr\,(X_1)^8+Z^2\,tr\,\left[ 8\,(X_1)^6(X_2)^2
+4\,(X_1)^4(X_2)^4+2\,(X_1)^2(X_2)^2(X_1)^2(X_2)^2\right]+\nonumber\\
  &+& Z^3\,tr\,\left[8\,(X_1)^4(X_2)^2(X_3)^2+8\,(X_1)^4X_2(X_3)^2
  X_2+8\,(X_1)^3(X_2)^2 X_1(X_3)^2+ \right. \nonumber\\
	&& \quad
	\left. +4\,(X_1)^2(X_2)^2(X_1)^2(X_3)^2\right]+\nonumber\\
	&+& Z^4\,tr\, \left[8\,(X_1)^2(X_2)^2 X_3(X_4)^2 X_3+
	4\, (X_1)^2 X_2X_3(X_4)^2 X_3X_2+2\, (X_1)^2(X_2)^2
	(X_3)^2(X_4)^2 \right] \qquad \nonumber\\ \frac{1}{N}
	tr\,A^{10} &=& Z\,tr\,(X_1)^{10}+Z^2\,tr\,10\,\left[
	(X_1)^8(X_2)^2+(X_1)^6(X_2)^4+(X_1)^4(X_2)^2(X_1)^2(X_2)^2\right]
	+\qquad \qquad \qquad \nonumber\\ &+& Z^3\,tr \,\left[
	10\,(X_1)^6(X_2)^2(X_3)^2+ 10\,(X_1)^6X_2(X_3)^2X_2+
	10\,(X_1)^5(X_2)^2X_1(X_3)^2 + \right. \nonumber\\ && \quad
	+10\,(X_1)^4(X_2)^4(X_3)^2+10\,(X_1)^4(X_2)^2(X_1)^2(X_3)^2+10\,(X_1)^4(X_2)^2
	(X_3)^2(X_2)^2 +  \nonumber\\ && \quad
	+10\,(X_1)^4(X_2)^3(X_3)^2 X_2+10\,(X_1)^4 X_2(X_3)^2(X_2)^3+10\,
	(X_1)^2(X_2)^2(X_1)^2X_2)^2(X_3)^2+\qquad \qquad \qquad
	\nonumber\\ && \quad \left. +10\,(X_1)^2 (X_2)^2(X_1)^2 X_2(X_3)^2
	X_2+5 (X_1)^4X_2(X_3)^4X_2+5(X_1)^3(X_2)^2(X_1)^3(X_3)^2
	\right] +\nonumber\\ &+& Z^4\,tr \,10\,\left[(X_1)^4X_2
	X_3(X_4)^2X_3X_2+(X_1)^4(X_2)^2X_3(X_4)^2X_3+
	(X_1)^4X_2(X_3)^2X_2(X_4)^2+  \right. \qquad \qquad \qquad
	\nonumber\\ && \quad +(X_1)^4 X_2(X_3)^2(X_4)^2X_2+
	(X_1)^4(X_2)^2(X_3)^2(X_4)^2+ (X_1)^3(X_2)^2 (X_3)^2
	X_1(X_4)^2   +\nonumber\\ && \quad +(X_1)^3 (X_2)^2
	X_1(X_3)^2(X_4)^2+(X_1)^3(X_2)^2 X_1 X_3 (X_4)^2 X_3+ (X_1)^3
	X_2(X_3)^2 X_2 X_1(X_4)^2 +\nonumber\\ && \quad  +(X_1)^2 (X_2)^2
	(X_3)^2X_2(X_4)^2X_2+ (X_1)^2(X_2)^2(X_1)^2X_3(X_4)^2X_3+\qquad
	\qquad \qquad  \nonumber\\ && \quad
	+\left.(X_1)^2(X_2)^2(X_1)^2(X_3)^2(X_4)^2\right]+\nonumber\\
	&+& Z^5\,tr \,\left[ 2(X_1)^2(X_2)^2(X_3)^2(X_4)^2(X_5)^2+
	5\,(X_1)^2(X_2)^2 X_3 (X_4)^2(X_5)^2X_3+\right.\qquad \qquad
	\qquad \nonumber\\ && \quad  + 5\,(X_1)^2 X_2X_3X_4(X_5)^2
	X_4X_3X_2+10\, (X_1)^2(X_2)^2(X_3)^2 X_4 (X_5)^2 X_4 +\nonumber\\
	&& \quad \left. + 10\,(X_1)^2 X_2(X_3)^2 X_4(X_5)^2 X_4X_2+10\,
	(X_1)^2(X_2)^2X_3X_4(X_5)^2X_4X_3 \right] \nonumber\\
	\frac{1}{N} tr\,A^{12} &=& Z\,tr\,(X_1)^{12}+Z^2\,tr\,2\,\left[ 6
	(X_1)^{10}(X_2)^2+6(X_1)^8(X_2)^4+3(X_1)^6(X_2)^6+\right.\nonumber\\
	&& \quad
	+6(X_1)^6(X_2)^2(X_1)^2(X_2)^2+6(X_1)^4(X_2)^4(X_1)^2(X_2)^2+3(X_1)^4(X_2)^2(X_1)^4(X_2)^2+\qquad
	\qquad \qquad \nonumber\\ && \quad \left.
	+(X_1)^2(X_2)^2(X_1)^2(X_2)^2(X_1)^2(X_2)^2\right]+\nonumber\\
	&&  +Z^3 \,tr \Bigg[ 12 \left[
	(X_1)^8(X_2)^2(X_3)^2+(X_1)^8X_2(X_3)^2X_2+(X_1)^7(X_2)^2X_1(X_3)^2+
	\right. \qquad \qquad \qquad \nonumber\\ && \quad
	+(X_1)^6(X_2)^4(X_3)^2+(X_1)^6(X_2)^2(X_3)^4+(X_1)^6X_2(X_3)^2(X_2)^3+(X_1)^6(X_2)^3(X_3)^2X_2
	+ \nonumber\\ && \quad
	+(X_1)^6X_2(X_3)^4X_2+(X_1)^6(X_2)^2(X_3)^2(X_2)^2+(X_1)^6(X_2)^2(X_1)^2(X_3)^2+\nonumber\\
		&& \quad \left. +(X_1)^5(X_2)^4X_1(X_3)^2+
 (X_1)^5(X_2)^2X_1(X_3)^4+(X_1)^5(X_2)^2(X_1)^3(X_3)^2 \right]+\nonumber\\
&& \quad +4 (X_1)^4(X_2)^4(X_3)^4+12 \left[
(X_1)^4(X_2)^4(X_1)^2(X_3)^2+(X_1)^4(X_2)^4(X_3)^2(X_2)^2+\right.
\nonumber\\ && \quad \left.+(X_1)^4(X_2)^3(X_3)^2(X_2)^3+
(X_1)^4(X_2)^3(X_3)^4 X_2 \right]+ 6 \left[
(X_1)^4(X_2)^2(X_1)^4(X_3)^2+ \right.\qquad \qquad \qquad
\nonumber\\ && \quad \left. +(X_1)^4(X_2)^2(X_3)^4(X_2)^2 \right]+12
\left[ (X_1)^4(X_2)^2(X_3)^2(X_1)^2(X_3)^2+\right. \nonumber\\ &&\quad
+(X_1)^4(X_2)^2(X_3)^2(X_1)^2(X_2)^2+(X_1)^4(X_2)^2(X_3)^2(X_2)^2(X_3)^2+(X_1)^4(X_2)^2(X_1)^2(X_3)^2(X_2)^2+
\nonumber\\ && \quad +(X_1)^4(X_2)^2(X_1)^2(X_2)^2(X_3)^2 +
(X_1)^4(X_2)^2(X_1)^2 X_2 (X_3)^2 X_2+ (X_1)^4 X_2 (X_3)^2
X_2(X_1)^2(X_2)^2 +\qquad \qquad \qquad \nonumber\\ && \quad
+(X_1)^4(X_2)^2 X_1 (X_3)^2 X_1(X_2)^2 + (X_1)^4 X_2 (X_3)^2(X_2)^2(X_3)^2
X_2 +
	\qquad \qquad \nonumber\\ && \quad +(X_1)^3(X_2)^2 X_1 (X_3)^2
	(X_1)^2(X_3)^2 +(X_1)^3(X_2)^2 (X_1)^2 (X_2)^2 X_1(X_3)^2 +\qquad
	\qquad \nonumber\\
		&& \quad \left. +(X_1)^2(X_2)^2 (X_1)^2 (X_2)^2
		(X_1)^2(X_3)^2 \right]+6\, (X_1)^2(X_2)^2 (X_1)^2 (X_3)^2
		(X_2)^2(X_3)^2+\qquad \qquad \nonumber\\ && \quad +2\,
		(X_1)^2(X_2)^2 (X_3)^2 (X_1)^2 (X_2)^2(X_3)^2 +3\, (X_1)^2
		X_2 (X_3)^2 X_2 (X_1)^2 X_2(X_3)^2 X_2 \Bigg]+\qquad
		\qquad \nonumber\\
	&&  +Z^4 \,tr \Bigg[	12\,\Bigg( X_1^6X_2^2X_3^2X_4^2     +
	   X_1^6X_2 X_3^2X_4^2X_2    +	X_1^6X_2X_3^2X_2X_4^2 +
	   X_1^6 X_2X_3X_4^2X_3X_2    \nonumber\\
&+&	X_1^6X_2^2X_3X_4^2X_3	+  X_1^5X_2^2X_3^2X_1X_4^2 +
X_1^5X_2^2X_1X_3^2X_4^2  + X_1^5X_2X_3^2X_2X_1X_4^2\nonumber\\ &+&
X_1^5X_2^2X_1X_3X_4^2X_3  + X_1^4X_2^4X_3^2X_4^2 +  X_1^4X_2^4 X_3 X_4^2
X_3 +  X_1^4 X_2 X_3^4 X_2 X_4^2\nonumber\\
 & +&	 X_1^4X_2X_3^4X_4^2 X_2 +  X_1^4 X_2^3 X_3^2 X_2 X_4^2 +
 X_1^4X_2^2X_3^3X_4^2 X_3 + X_1^4 X_2 X_3^2 X_4^2 X_2^3   \nonumber\\
 &+& X_1^4 X_2^2  X_3 X_4^2 X_3^3 + X_1^4X_2^3 X_3^2 X_4^2X_2 +  X_1^4
 X_2 X_3^2 X_2^3 X_4^2 + X_1^4 X_2X_3 X_4^2X_3X_2^3\nonumber\\ &+&
 X_1^4X_2X_3^3X_4^2X_3X_2  +  X_1^4 X_2X_3 X_4^2X_3^3 X_2  +  X_1^4
 X_2^3 X_3 X_4^2X_3X_2	+ X_1^4 X_2^2 X_1^2X_3^2 X_4^2 \nonumber\\
&+&  X_1^4 X_2^2 X_3^2 X_1^2 X_4^2 +  X_1^4 X_2^2 X_3^2 X_4^2 X_3^2+
X_1^4 X_2^2 X_3^2 X_2^2 X_4^2 +  X_1^4 X_2^2 X_3^2 X_4^2 X_2^2 \nonumber\\
 &+& X_1^4 X_2^2 X_1^2 X_3 X_4^2 X_3+X_1^4X_2X_3^2 X_2 X_1^2 X_4^2 +X_1^4
 X_2 X_3^2 X_2^2 X_4^2 X_2 + X_1^4 X_2 X_3^2 X_2 X_4^2 X_2^2\nonumber\\
&+&  X_1^4X_2^2 X_3^2 X_2 X_4^2 X_2+X_1^4 X_2^2 X_1 X_3^2 X_1 X_4^2+X_1^4
X_2^2 X_3 X_4^2 X_3 X_2^2 +X_1^4 X_2 X_3^2 X_4^2 X_3^2 X_2\nonumber\\
 &+& 6\, \Bigg(   X_1^4X_2^2X_3^4X_4^2+X_1^4X_2X_3 X_4^4 X_3 X_2 +
 X_1^3 X_2 X_3^2 X_2^3 X_1 X_4^2+ X_1^3 X_2^2 X_1 X_3^3  X_4^2X_3\Bigg)
 \nonumber\\
&+& 12\, \Bigg( X_1^3X_2^2 X_1^3 X_3^2 X_4^2+ X_1^3 X_2^2 X_1^3 X_3
X_4^2 X_3+ X_1^3 X_2^2 X_1^2 X_3^2 X_1 X_4^2+ X_1^3 X_2^2 X_1 X_3^2
X_4^2 X_3^2 \nonumber\\ &+& X_1^3X_2^2 X_1 X_3^2 X_1^2 X_4^2+ X_1^3X_2^2
X_3^2 X_2^2 X_1 X_4^2 + X_1^3 X_2^3 X_3^2 X_2 X_1 X_4^2+X_1^2 X_2^2 X_1^2
X_2^2 X_3^2 X_4^2 \nonumber\\ &+&  X_1^2 X_2^2 X_1^2 X_3^2 X_2^2 X_4^2+
 X_1^2 X_2^2 X_3^2 X_1^2 X_3 X_4^2 X_3 +X_1^2 X_2^2 X_3^2 X_1^2 X_2
 X_4^2 X_2 + X_1^2 X_2^2 X_1^2 X_3^2 X_2 X_4^2 X_2 \nonumber\\
&+& X_1^2 X_2^2 X_3^2 X_2^2 X_3 X_4^2 X_3+	X_1^2 X_2^2 X_1^2 X_2^2
X_3 X_4^2 X_3+	X_1^2 X_2^2 X_1^2 X_2 X_3^2 X_4^2 X_2 \nonumber\\ &+&
X_1^2 X_2^2  X_1^2 X_2 X_3 X_4^2 X_3 X_2+ X_1^2 X_2^2 X_1 X_3^2 X_1 X_2
X_4^2 X_2\Bigg)\nonumber\\ &+& 6\,\Bigg( X_1^2 X_2^2 X_1^2 X_3^2 X_4^2
X_3^2+ X_1^2 X_2^2X_3^2X_1^2 X_2^2 X_4^2 +  X_1^2 X_2 X_3^2 X_2 X_1^2
X_2 X_4^2 X_2 \Bigg) \nonumber\\ &+&   4\,  X_1^2 X_2^2 X_1^2 X_3^2
X_1^2 X_4^2 \Bigg]\nonumber\\ &&  +Z^5 \,tr
 \,\Bigg[12\, \Bigg(  X_1^4X_2^2X_3^2X_4^2 X_5^2  + X_1^4X_2^2 X_3^2
 X_4 X_5^2 X_4	+    X_1^4 X_2^2 X_3 X_4^2 X_3 X_5^2 \nonumber\\ &&
 +X_1^4 X_2 X_3^2 X_2 X_4^2 X_5^2   + X_1^4 X_2 X_3^2  X_4^2 X_5^2 X_2	+
 X_1^4 X_2^2 X_3 X_4^2 X_5^2 X_3\nonumber\\
&& +X_1^4 X_2 X_3^2 X_4^2 X_2 X_5^2 +  X_1^4 X_2 X_3 X_4^2 X_3 X_2 X_5^2 +
X_1^4 X_2^2 X_3 X_4 X_5^2 X_4 X_3 \nonumber\\
	&& +X_1^4 X_2 X_3 X_4^2 X_5^2 X_3 X_2 +
	 X_1^4 X_2 X_3^2 X_4 X_5^2 X_4 X_2 + X_1^4 X_2 X_3 X_4^2 X_3
	 X_5^2 X_2 \nonumber\\
	&& +X_1^4 X_2 X_3^2 X_2 X_4 X_5^2 X_4  + X_1^4 X_2 X_3 X_4 X_5^2
	X_4 X_3 X_2 \nonumber\\ &&  +X_1^3 X_2^2 X_3^2 X_4^2 X_1 X_5^2 +
	X_1^3 X_2^2 X_1 X_3^2 X_4^2 X_5^2 + X_1^3 X_2^2 X_3^2 X_1 X_4^2
	X_5^2\nonumber\\ &&  +X_1^3 X_2 X_3^2 X_2 X_1 X_4^2 X_5^2 +
	  X_1^3 X_2^2 X_3^2 X_1 X_4 X_5^2 X_4  + X_1^3 X_2^2 X_3 X_4^2
	  X_3 X_1 X_5^2 \nonumber\\
	&& +X_1^3 X_2 X_3^2 X_2 X_4^2 X_1 X_5^2	+ X_1^3 X_2^2 X_1 X_3^2
	X_4 X_5^2 X_4 + X_1^3 X_2^2 X_1 X_3 X_4^2 X_3 X_5^2 \nonumber\\
	 &&  +X_1^3 X_2^2 X_1 X_3 X_4^2 X_5^2  X_3  + X_1^3 X_2 X_3^2
	 X_4^2 X_2 X_1 X_5^2 + X_1^3 X_2 X_3 X_4^2 X_3 X_2 X_1 X_5^2
	 \nonumber\\
  && +X_1^3 X_2^2 X_1 X_3 X_4 X_5^2 X_4 X_3	+ X_1^3 X_2 X_3^2 X_2
  X_1 X_4 X_5^2 X_4  \nonumber\\
	&& +X_1^2 X_2^2 X_1^2 X_3^2 X_4^2 X_5^2 + X_1^2 X_2^2 X_1^2 X_3^2
	X_4 X_5^2 X_4 +X_1^2 X_2^2 X_3^3 X_1 X_4^2 X_1 X_5^2\nonumber\\
&&	+X_1^2 X_2^2 X_1 X_3^2 X_1 X_4^2 X_5^2 +X_1^2 X_2^2 X_1^2 X_3
X_4^2 X_3 X_5^2 +X_1^2 X_2^2 X_3^2 X_1^2 X_4 X_5^2 X_4 \nonumber\\
	&&	 +X_1^2 X_2^2 X_1^2 X_3 X_4 X_5^2 X_4 X_3 + X_1^2 X_2^2
	X_1 X_3 X_4^2 X_3 X_1 X_5^2 +X_1^2 X_2^2 X_1^2 X_3 X_4^2 X_5^2
	X_3 \nonumber\\
 && +X_1^2 X_2 X_3^2 X_2 X_1 X_4^2 X_1 X_5^2 + X_1^2 X_2^2 X_1 X_3^2 X_1
 X_4 X_5^2 X_4 + X_1^2 X_2^2 X_1 X_3^2 X_4^2 X_1 X_5^2\Bigg) \nonumber\\
&& +6 \, X_1^2 X_2^2 X_3^2 X_1^2 X_4^2 X_5^2 +
	  +6\, X_1^2 X_2 X_3^2 X_2 X_1^2 X_4 X_5^2 X_4+3\, X_1^2 X_2
	  X_3^2 X_2 X_4^2 X_2 X_5^2 X_2 \Bigg]+ \nonumber\\
		&+& Z^6\,tr \,\Bigg[ 12 \Bigg( X_1^2 X_2^2 X_3^2 X_4^2
		X_5 X_6^2 X_5+X_1^2 X_2^2 X_3^2 X_4 X_5^2 X_6^2 X_4+X_1^2
		X_2^2 X_3^2 X_4 X_5 X_6^2 X_5 X_4 \nonumber\\ && X_1^2
		X_2^2 X_3 X_4^2 X_5 X_6^2 X_5 X_3+X_1^2 X_2^2 X_3 X_4^2
		X_3 X_5 X_6^2 X_5+X_1^2 X_2^2 X_3 X_4 X_5^2 X_4 X_6^2
		X_3 \nonumber\\ && X_1^2 X_2^2 X_3 X_4 X_5 X_6^2 X_5
		X_4 X_3+X_1^2 X_2 X_3^2 X_4 X_5 X_6^2 X_5 X_4 X_2 +X_1^2
		X_2 X_3^2 X_2 X_4 X_5 X_6^2 X_5 X_4
			\Bigg)+\nonumber\\
		&& 6 \Bigg( X_1^2 X_2^2 X_3 X_4 X_5^2 X_6^2 X_4 X_3+
		X_1^2 X_2 X_3^2 X_2 X_4^2 X_5 X_6^2 X_5+ X_1^2 X_2 X_3
		X_4 X_5 X_6^2 X_5 X_4 X_3 X_2 \Bigg)+ \nonumber\\ &&
		4\, X_1^2 X_2 X_3 X_4^2  X_3 X_5 X_6^2 X_5 X_2+ 2\,
		X_1^2 X_2^2 X_3^2 X_4^2 X_5^2 X_6^2 \Bigg]
\nonumber\\
	\label{A.1}
\end{eqnarray} \normalsize

	This table of products of blocks $X_j$ allows to evaluate
	analytically the limiting ($N \to \infty$) spectral moments for
	any sparse ensemble of real symmetric random block matrices
	$Nd \times Nd$, where the random $d \times d$ real symmetric
	blocks are i.i.d., and the architecture of the Adjacency matrix
	corresponds to the Erd\"os-Renyi  random graph, with average
	degree $Z$.\\

	Actually, it is possible to obtain also the limiting spectral
	moments of  real symmetric random block matrices $Nd \times
	Nd$, where the random $d \times d$ real symmetric  blocks are
	i.i.d., and the architecture of the Adjacency matrix corresponds
	to regular random graphs, with fixed degree $Z$.  This requires
	only the determination of the $Z-$dependent weight,  as  we show
	with one example.\\

	Erd\"os-Renyi  random graph : \begin{eqnarray} \frac{1}{N}
	tr\,A^8 &=& Z\,tr\,X_1^8+Z^2\,tr\,\left[ 8\,X_1^6 X_2^2
	+4\,X_1^4 X_2^4+2\,X_1^2X_2^2X_1^2X_2^2\right]+\nonumber\\
  &+& Z^3\,tr\,\left[8\,X_1^4X_2^2X_3^2+8\,X_1^4X_2X_3^2
  X_2+8\,X_1^3X_2^2 X_1X_3^2+ \right. \nonumber\\
	&& \quad
	\left. +4\,X_1^2X_2^2X_1^2X_3^2\right]+\nonumber\\ &+&
	Z^4\,tr\, \left[8\,X_1^2X_2^2 X_3X_4^2 X_3+ 4\, X_1^2
	X_2X_3X_4^2 X_3X_2+2\, X_1^2X_2^2 X_3^2X_4^2 \right]
	\nonumber \end{eqnarray}

Random regular graph :
	\begin{eqnarray}
\frac{1}{N} \texttt{tr} A^8 &=& Z\,\texttt{tr} \,X_1^8+
Z(Z-1)\,\texttt{tr} \bigg( 8 \,X_1^6 X_2^2+4 \,X_1^4 X_2^4	+2\,X_1^2
X_2^2 X_1^2 X_2^2 \bigg) +\nonumber\\	&&Z(Z-1)(Z-2)\,\texttt{tr} \bigg(8
\, X_1^4 X_2^2 X_3^2+4 \, X_1^2 X_2^2 X_1^2 X_3^2 \bigg) +\nonumber\\
   &&Z(Z-1)^2 \,\texttt{tr}\bigg( 8  \,X_1^4 X_2 X_3^2 X_2+8 \, X_1^3
   X_2^2 X_1 X_3^2 \bigg)+ \nonumber\\
	  &&  Z(Z-1)^2 (Z-2)\,8\, \texttt{tr}\, X_1^2 X_2^2 X_3 X_4^2
	  X_3+	Z(Z-1)^3\, 4 \,\texttt{tr} \, X_1^2 X_2 X_3 X_4^2 X_3
	  X_2\nonumber\\
		&&   Z(Z-1)(Z-2)(Z-3)\,2\, \texttt{tr} \,X_1^2 X_2^2
		X_3^2 X_4^2
\nonumber
  \end{eqnarray}

 \vskip 1.2cm

Table 2. Limiting moments (that is $N \to \infty $) of the Adjacency
matrix, evaluation of the averages for finite $d$.  \vskip 0.8 cm
\begin{eqnarray}
 \frac{1}{Nd}< tr\,A^2> &=& Z(d+1) \, \sigma^2 \nonumber\\
\frac{1}{Nd} <tr\,A^4> &=&  \Bigg[ Z(2 d^2+5d+5)+Z^2 2(d+1)^2\Bigg]
\sigma^4 \nonumber\\ \frac{1}{Nd} <tr\,A^6> &=&  \Bigg[ Z(5d^3+22 d^2+52
d+41)+Z^2 6(d+1)(2d^2+5d+5)+Z^3 5(d+1)^3\Bigg]\sigma^6 \nonumber\\
\frac{1}{Nd} <tr\,A^8> &=&  \Bigg[ Z(14 d^4+93d^3+374 d^2+690 d+509)+
Z^2 \left[ 8(d+1)(5d^3+22d^2+52d+41)+\right.\nonumber\\
   &+&	  \left.  4(2d^2+5d+5)^2+
 6(d^4+5d^3+13 d^2+18d+11) \right] +\nonumber\\
&+& Z^3   28(d+1)^2(2d^2+5d+5)	+ Z^4 \,14(d+1)^4 \Bigg]\sigma^8
\nonumber\\
 \label{A.2}
\end{eqnarray}

\vskip	1cm
 Table 3.  Spectral moments $f_n$ in terms of free cumulants $a_n$.
\vskip 0.4 cm \begin{eqnarray} f_1 &=& a_1  \nonumber\\ f_2 &=&
a_2+a_1^2 \nonumber\\ f_3 &=& a_3+3a_1a_2+a_1^3 \nonumber\\ f_4 &=&
a_4+4a_1 a_3+2 a_2^2+6a_1^2 a_2+a_1^4 \nonumber\\ f_5 &=& a_5+5a_1a_4
+5 a_2 a_3+10 a_1^2 a_3 +10 a_1 a_2^2+10 a_1^3 a_2+a_1^5\nonumber\\
f_6 &=& a_6+6 a_1 a_5+6 a_2 a_4+15 a_1^2 a_4+3 a_3^2+30 a_1 a_2 a_3+20
a_1^3 a_3+5 a_2^3+30 a_1^2 a_2^2+15 a_1^4 a_2+a_1^6 \nonumber\\ f_7 &=&
a_7+7 a_1 a_6+7 a_2 a_5+21 a_1^2 a_5+7 a_3 a_4+ 42 a_1 a_2 a_4+35 a_1^3
a_4+21 a_1 a_3^2+21 a_2^2 a_3+\nonumber\\
 &&\quad 105 a_1^2 a_2 a_3+35 a_1^4 a_3+35 a_1 a_2^3+70 a_1^3 a_2^2+21
 a_1^5 a_2 +a_1^7 \nonumber\\
f_8 &=& a_8+ 8 a_1 a_7+8 a_2 a_6+28 a_1^2 a_6+8 a_3 a_5+56 a_1 a_2 a_5+56
a_1^3 a_5+4 a_4^2+56 a_1 a_3 a_4+28 a_2^2 a_4+\nonumber\\
  && \quad  168 a_1^2 a_2 a_4+70 a_1^4 a_4+28  a_2 a_3^2+84 a_1^2 a_3^2
  +168 a_1 a_2^2 a_3+280 a_1^3 a_2 a_3+56 a_1^5 a_3+14 a_2^4+\nonumber\\
	&& \quad	140 a_1^2 a_2^3+140  a_1^4 a_2^2+28 a_1^6 a_2+a_1^8
\nonumber \end{eqnarray} \begin{eqnarray} f_n=a_n+\sum_{k=2}^n\frac{1}{k}
\left( \begin{array}{cc} n \\ k-1 \end{array} \right) \sum_{Q_k} a_{q_1}
a_{q_2} ..a_{q_k} \qquad \label{m.1} \end{eqnarray} where $Q_k$ is
the set of integers : $$ Q_k=\Bigg\{ (q_1,..,q_k)\in N^k \, \Bigg| \,
\sum_{i=1}^kq_i=n \Bigg\} $$

Notice that, if all $a_j=1$ then $f_n= C_n=\frac{ 1}{n}\left(
\begin{array}{cc} 2n\\n-1 \end{array}\right)$.\\

The inverse relation, irreducible elements $a_j$ in term of spectral
moments is the following.

\begin{eqnarray} a_1 &=& f_1 \nonumber\\ a_2 &=& f_2-f_1^2 \nonumber\\ a_3
&=& f_3-3 f_1 f_2+2 f_1^3 \nonumber\\ a_4 &=& f_4-4 f_1 f_3+10 f_1^2 f_2-2
f_2^2 -5 f_1^4 \nonumber\\ a_5 &=& f_5-5 f_1 f_4-5 f_2 f_3+15 f_1^2 f_3+15
f_1 f_2^2-35 f_1^3 f_2+14 f_1^5 \nonumber\\ a_6 &=& f_6-6f_1f_5-6f_2
f_4+21 f_1^2 f_4-3 f_3^2+42 f_1f_2 f_3-56 f_1^3 f_3+7 f_2^3-84 f_1^2
f_2^2+\nonumber\\ && \quad 126 f_1^4f_2-42 f_1^6 \nonumber\\ a_7 &=&
f_7-7 f_1f_6-7 f_2 f_5-7 f_3 f_4+28 f_1^2 f_5+56f_1f_2f_4-84 f_1^3
f_4+28 f_1 f_3^2+\nonumber\\ &&\quad 28 f_2^2f_3-252 f_1^2f_2f_3+210
f_1^4f_3-84f_1f_2^3+420 f_1^3f_2^2-462f_1^5f_2+132 f_1^7 \nonumber\\ a_8
&=&  f_8- 8 f_1f_7-8 f_2 f_6-8 f_3 f_5+36 f_1^2 f_6+72 f_1f_2f_5 -120
f_1^3 f_5-4 f_4^2+72 f_1f_3 f_4+ \nonumber\\ && \quad 36 f_2^2 f_4-360
f_1^2 f_2 f_4+330 f_1^4f_4+36 f_2 f_3^2-180 f_1^2 f_3^2-360f_1f_2^2 f_3+
\nonumber\\ && \quad 1320 f_1^3 f_2 f_3-792 f_1^5f_3-30 f_2^4+660 f_1^2
f_2^3-1980 f_1^4f_2^2+1716 f_1^6f_2-429 f_1^8 \nonumber \end{eqnarray}
\begin{eqnarray} a_n=f_n+\sum_{k=2}^n \frac{ (-1)^{k-1}}{k} \left(
\begin{array}{cc} n+k-2 \\ k-1 \end{array} \right) \sum_{Q_k} f_{q_1}
f_{q_2}..f_{q_k}\qquad \label{m.2} \end{eqnarray}

Equations (\ref{m.1}), (\ref{m.2}) appear in \cite{motte}. Summations
over the set of non-crossing partitions are in \cite{spei}, \cite{nica}.

\vskip	1cm
 Table 4. Limiting Moments of the product of two random matrices
 \cite{bai2007}.
\vskip 0.4 cm \begin{eqnarray} \beta_2 &=& 1  \nonumber\\ \beta_4
&=& 2+2\,y \nonumber\\ \beta_{6} &=& 5+12\,y+5\, y^2 \nonumber\\
\beta_{8} &=& 14+56 \,y+ 56\, y^2+14\, y^3 \nonumber\\ \beta_{10} &=&
42 +240\,y+405\,y^2+240\,y^3+42\,y^4 \nonumber\\ \beta_{12} &=& 132 +990
\,y+ 2420 \,y^2+2420\,y^3+  +990\,y^4  +132 \, y^5
 \nonumber
\end{eqnarray}

\begin{eqnarray}
 \beta_k =\Bigg\{ \begin{array}{ccccccccc} \sum_{j=1}^{k/2} \left(
 \begin{array}{ccc} k\\ j-1 \end{array}\right)\left( \begin{array}{ccc}
 k\\ k/2-j \end{array}\right)\frac{2}{k} \,y^{j-1} \qquad \texttt{if $k$
 is even},\\ 0 \qquad\qquad \texttt{if $k$ is odd}.
\end{array} \qquad \label{A.6} \end{eqnarray}


 \end{document}